\documentclass{article}
\usepackage{spconf,extra_styles,amsmath,graphicx}
\usepackage{mathtools}
\usepackage{bbm}
\usepackage{makecell} 
\usepackage{threeparttable}
\usepackage{pifont}
\usepackage{multirow}
\usepackage{booktabs}
\usepackage[pagebackref,breaklinks,colorlinks]{hyperref}

\title{Seeing Through the Conversation: \\Audio-Visual Speech Separation based on Diffusion Model}
\name{Suyeon Lee$^*$, Chaeyoung Jung$^*$, Youngjoon Jang, Jaehun Kim, Joon Son Chung\thanks{$^*$ These authors contributed equally.}}
\address{Korea Advanced Institute of Science and Technology, South Korea}

\begin{document}
\ninept
\maketitle
\begin{abstract}
The objective of this work is to extract target speaker's voice from a mixture of voices using visual cues. 
Existing works on audio-visual speech separation have demonstrated their performance with promising intelligibility, but maintaining naturalness remains a challenge.
To address this issue, we propose AVDiffuSS, an audio-visual speech separation model based on a diffusion mechanism known for its capability in generating natural samples. For an effective fusion of the two modalities for diffusion, we also propose a cross-attention-based feature fusion mechanism.
This mechanism is specifically tailored for the speech domain to integrate the phonetic information from audio-visual correspondence in speech generation.
In this way, the fusion process maintains the high temporal resolution of the features, without excessive computational requirements.
We demonstrate that the proposed framework achieves state-of-the-art results on two benchmarks, including VoxCeleb2 and LRS3, producing speech with notably better naturalness.
Project page with demo: \url{https://mm.kaist.ac.kr/projects/avdiffuss/}

\end{abstract}
\section{Introduction}
\label{sec:intro}

While significant advancements in audio-only speech recognition and separation techniques have been witnessed recently, challenges remain in understanding a speech from an individual amidst overlapping sounds. 
In real-world situations, conversations are often intertwined with other voices or disturbed by a cacophony of noises. 
Elimination of such disturbances is particularly important in settings like meetings, where one has to focus on the speech of single individual.
Humans excel at guiding their attention to a sound source of interest in such environments, naturally de-emphasizing other sounds.
On the other hand, when auditory cues contradict visual cues from the speaker's face, speech sounds are frequently misinterpreted by humans~\cite{mcgurk1976hearing}, highlighting the importance of visual modality in human's understanding of spoken communications.

\begin{figure}[!t]
    \centering
    \includegraphics[width=1.0\linewidth]{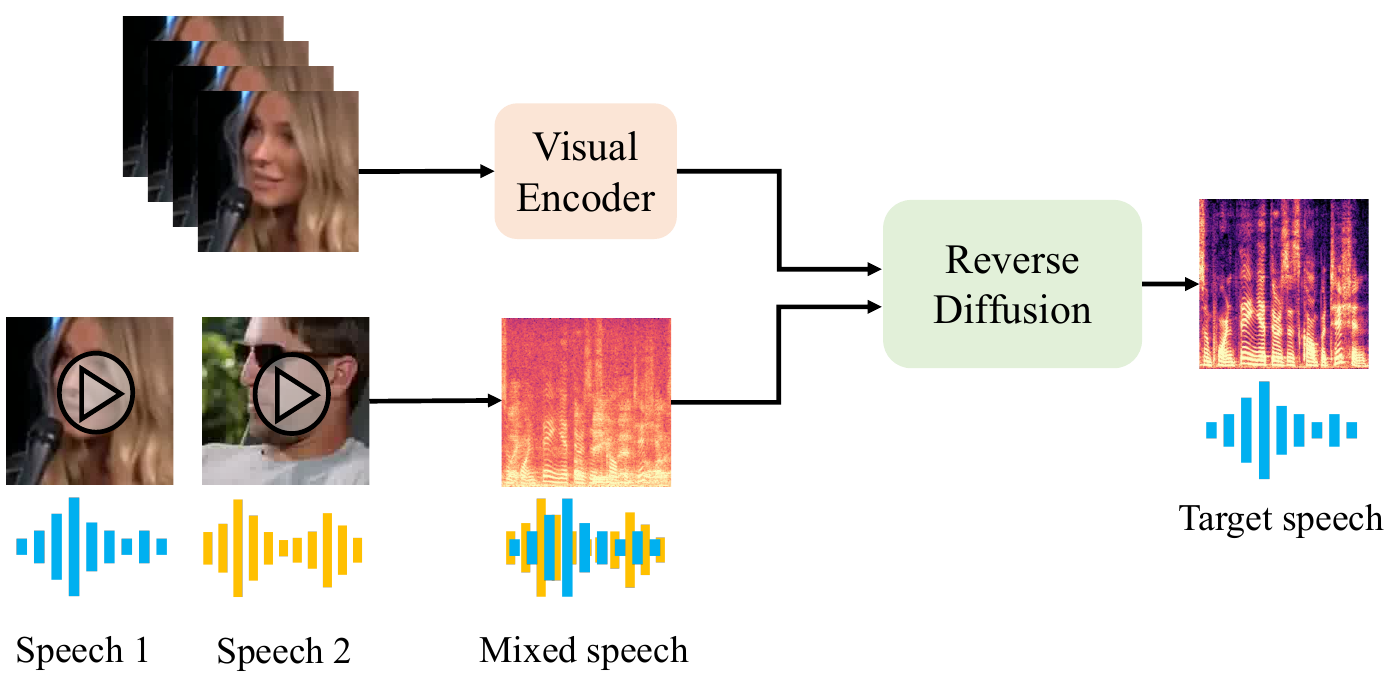}
    \vspace{-2mm}
    \caption{Audio-visual speech separation based on diffusion model pipeline. Mixed speech and target speaker's face crop images are inputs and target speaker's speech is extracted through a reverse diffusion process.}
    \vspace{-2mm}
    \label{fig1}
\end{figure}

Audio-Visual Speech Separation (AVSS) aims to emulate this human capacity, aiming to distinguish each voice from a collective soundscape using the visual information.
Beyond enhancing the auditory intelligibility for listeners, this technique can also serve as a pre-processing step for various speech-related tasks, including cascaded speech recognition~\cite{c29,tan2020audio} and speaker diarisation~\cite{wuerkaixi2022dyvise}.
In consequence, there has been significant advances in the field of audio-visual speech separation, driven by the accessibility of multi-modal datasets and high performance computing. 
Early works~\cite{c29, looking} have proposed to combine visual and audio features to distinguish the speech of the target speaker in complex and noisy environments.
A noteworthy finding in their research is that leveraging the visual modality effectively addresses the label permutation problem, which arises from the challenge of assigning a proper ground truth to the predicted output during training. 
More recently, VisualVoice~\cite{visualvoice} utilizes both lip motions and facial attributes (e.g. gender, age, and nationality) as conditions to specify the target speaker. 
Thus, it is reported that leveraging lip movements is effective for aligning auditory and visual information to extract phonetic information, and incorporating facial attributes aids in distinguishing target speakers using their identity cues. 

With the advancements in deep learning, there has been successful applications of generative models in AVSS field. 
Generative AVSS models~\cite{yang2022audio,nguyen2021deep} can produce realistic samples by learning the mapping from latent space to clean speech distribution.
Although these approaches have demonstrated successful performance, they face difficulties in generating diverse samples and exact data estimation, frequently producing speech with undesirable artifacts. 
This indicates the need for generating samples that sound more natural to humans. 
In response to this, we take advantage of the natural sample generation capabilities of the diffusion model. 
The diffusion model is known for its potential in generating diverse and natural samples across various domains~\cite{latentdif, storm,popov2021grad,sgmse+, ho2020denoising}, including the audio-only speech separation~\cite{Diffsep,lutati2023separate}.

\begin{figure*}[!t]
  \centering
  \includegraphics[width=1.0\linewidth]{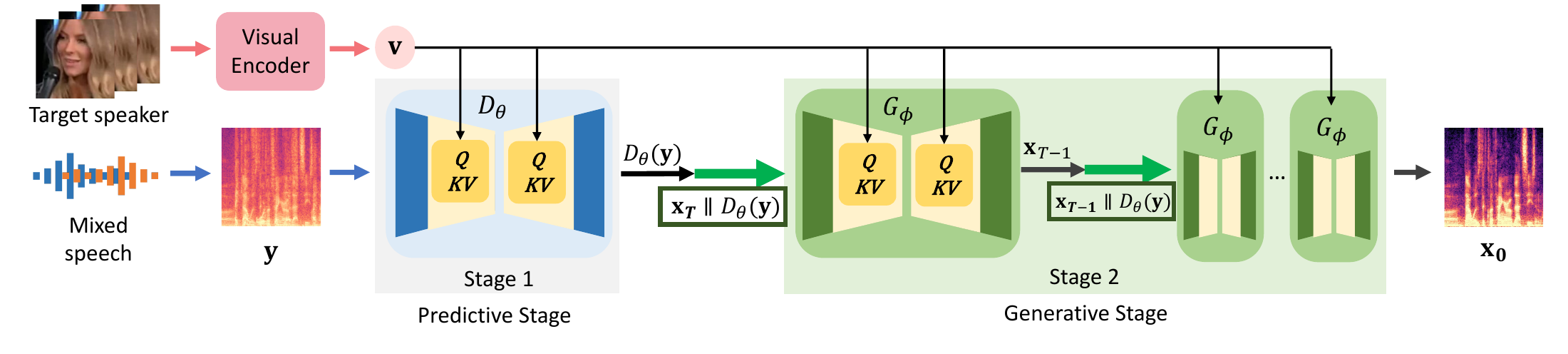}
  \vspace{-2mm}
  \caption{Model architecture of AVDiffuSS. Face-cropped images of the target speaker are fed to the visual encoder to obtain the visual embedding $\mathbf{v}$. A mixture of two speech signals is transformed into a spectrogram $\mathbf{y}$ by STFT, and it goes into two stages: 1) a predictive stage, and 2) a generative stage. Each stage consists of U-net architecture. For the input of the generative stage, the output of first stage $D_\theta(\mathbf{y})$ is concatenated with the $\mathbf{x_T}$, which is sampled from $\mathcal{N}(D_{\theta}(\mathbf{y}),\,\sigma^{2}\mathbf{I})\,$. In the next stage, reverse diffusion is repeated for $T$ steps. At the same time, $\mathbf{v}$ is utilized as a key and value and audio feature from U-net is used as a query, in the cross-attention modules of both stages.}
  \vspace{-2mm}
  \label{fig2}
\end{figure*}

In this work, we propose a diffusion-based AVSS model called \texttt{AVDiffuSS} that reconstructs both natural and intelligible utterances.
To effectively incorporate visual information in the separation process, \cite{latentdif} leverages a feature fusion mechanism. However, its fusion strategy based on feature compression is not appropriate in speech domain where time resolution must be preserved.
To mitigate this issue, we propose a task-specific feature fusion mechanism for visual gudiance.
The proposed cross-attention layer does not require any encoder and decoder module, eliminating the risk of performance bottleneck.
The layer supports a frequency-domain compression of audio features for reducing memory footprint, without time-domain compression.

Our contributions consist of the following:
(1) To the best of our knowledge, we are the first to introduce an audio-visual speech separation based on diffusion model, capable of reconstructing both natural and intelligible speech.
(2) With the help of the proposed compressing strategy, we successfully mitigate the excessive computational overheads and make our model suitable in speech domain.
(3) With various experiments, we demonstrate that the proposed method attains state-of-the-art results on two widely-used benchmark datasets.

\section{Method}
\label{sec:method}
As illustrated in~\Fref{fig2}, our framework comprises two main stages: the predictive stage and the generative stage. In the predictive stage, the model initially estimates the speech of the target speaker using visual semantics $v$ extracted by the visual encoder. The output of the predictive model, denoted as $D_{\theta}(y)$, is then fed into the generative stage, which employs a diffusion-based model. In this stage, the initial prediction is further enhanced through an iterative denoising process. Note that both stages improve audio-visual alignments by utilizing a task-specific cross-attention module, resulting in the generation of more natural samples.

\subsection{Visual Encoder}
\label{ssec:method_visenc}
The visual modality plays two pivotal roles for audio-visual speech separation: (1) synchronizing speech with lip movements to capture phonetic details, and (2) identifying the target speaker based on facial attributes, such as gender, age, and nationality. Taking inspiration from a study in active speaker detection~\cite{talknet}, a visual encoder capable of both preserving temporal dynamics and incorporating visual cues can be leveraged to achieve aforementioned objectives. We adopt the encoder architecture from~\cite{talknet}, which comprises a series of ResNet18 layers and a temporal convolutional network from~\cite{C30}. On top of those modules, a 1D convolution layer is attached to reduce the channel dimension. The encoder, as a result, outputs frame-level spatio-temporal features.

\subsection{Encoder-Decoder Free Conditioning by Cross-Attention}
\label{ssec:method_crossatt}

To effectively separate the desired speech by exploiting the visual modality, it is essential to maintain the temporal characteristics of both auditory and visual features throughout the fusion process. Based on this, we focus on the cross-attention mechanism, which enables the model to learn the correspondence between sequential information from the two different modalities.
Since cross-attention calculates correlations between different modalities by multiplication, it requires heavy computational costs. 
In response to this, we propose a feature fusion method using cross-attention, eliminating the need for a complex feature compression process involving encoder-decoder architecture.

The proposed feature fusion method is conducted in both predictive and generative stages. In each stage, we aim to acquire the correspondence between visual and audio embeddings. As we adopt the U-net architecture as the backbone of both stages, audio embedding can be represented as $\mathbf{e}_{a,i} \in \mathbbm{R}^{C_i \times T_i \times F_i}$. Here, $C_i$, $T_i$, and $F_i$ denote the number of channels, frame lengths, and frequency lengths, respectively, in the $i$-th U-net layer.
By applying frequency-axis pooling to the audio features, we obtain the pooled audio feature denoted as $\mathbf{\bar{e}}_{a,i} \in \mathbb{R}^{C_i \times T_i}$, which is used as the query, while the visual feature is employed as the key and value. The output of the cross-attention module is repeated $F_i$ times to recover the original shape of the input before averaging across frequencies.
Through this process, our model denoises undesired speech while enhancing the voice of the target speaker.

\subsection{Audio-Visual Speech Separation with Diffusion}
\label{ssec:method_ovrl}

\newpara{Diffusion model.}
\label{ssec:diff}
In diffusion process ${\{ x_\tau \}}^T_{\tau=0}$ , indexed by a continuous time variable $\tau \in [0,T]$ , data is perturbed with a noise during a forward process. The diffusion model~\cite{ho2020denoising,C28} learns to reverse this process to generate a clean data $\mathbf{x_0} \sim p_0$ from a noisy prior $\mathbf{x_T} \sim p_T$. 
Forward process is modeled using a Stochastic Differential Equation (SDE) as follows:

\begin{equation} 
\label{forward}
\mathrm{d} \mathbf{x_{\tau}} = f (\mathbf{x_{\tau}}, \tau) \mathrm{d} \tau+g(\tau) \mathrm{d} \mathbf{w},
\end{equation}

where $\mathbf{w}$ is a Brownian motion, $f(\mathbf{x_{\tau}},\tau) \vcentcolon= \gamma (\mathbf{y} - \mathbf{x_\tau})$ is a drift term following~\cite{storm,sgmse+} and $g(\tau)$ is a diffusion coefficient. The model learns to solve the reverse-time SDE, given as:

\begin{equation} 
\label{reverse}
\mathrm{d} \mathbf{x_{\tau}}=\left[f(\mathbf{x_{\tau}}, \tau)-g(\tau)^2 \nabla_{\mathbf{x_{\tau}}} \log p_{\tau}(\mathbf{x_{\tau}})\right] \mathrm{d} \tau+g(\tau) \mathrm{d} \overline{\mathbf{w}},
\end{equation} 

where $\overline{\mathbf{w}}$ denotes a Brownian motion for the reverse time flowing from $T$ to $0$. 
By training our generative stage on samples with denoising score matching objective, $G_\phi(\cdot)$ learns to transform a noisy data $\mathbf{x_T} \sim p_T$ into a clean sample $\mathbf{x_0} \sim p_0$.

\newpara{Training procedure.}
The architecture of both stages is U-net~\cite{ronneberger2015u} with cross-attention modules for the integration of audio and visual features~\cite{latentdif}, as shown in Fig.~\ref{fig2}. 
Audio input $\mathbf{y}$ denotes a complex-valued spectrogram of an overlapped speech, and each channel is occupied by the real and imaginary values respectively. 
In the predictive stage, the model $D_\theta(\cdot)$ is trained to directly separate the desired speech $\mathbf{x}$ from $\mathbf{y}$. The output of the first stage $D_\theta(\mathbf{y})$ is utilized to sample the initial condition $\mathbf{x_T}$ for the second stage as follows:

\begin{equation} 
\label{eq1}
\mathbf{x_T} \sim \mathcal{N}(D_{\theta}(\mathbf{y}),\,\sigma^{2}\mathbf{I})\,,
\end{equation}

where $\mathbf{x_T}$ can be simply obtained by adding a Gaussian noise to the initial prediction $D_\theta(\mathbf{y})$.
$D_\theta(\mathbf{y})$ is concatenated with $\mathbf{x_T}$ to construct the input for the generative stage and $\mathbf{x_{T-1}}$ results from the first step through $G_\phi(\cdot)$ as follows:

\begin{equation} 
\label{concat}
\mathbf{x_{T-1}} = G_\phi (\mathbf{x_T} \mathbin\Vert D_\theta(\mathbf{y})) .
\end{equation}

$\mathbf{x_{T-1}}$ is again fed to $G_\phi$, and this process is repeated for $T$ steps. Through this reverse-diffusion process, the target audio prediction $\mathbf{x_0}$ could be separated from $\mathbf{y}$.

\begin{table}
\centering
\resizebox{\columnwidth}{!}{%
\begin{tabular}{lccccc}
\toprule
Method   & A-V & Diff & PESQ   & ESTOI  & SI-SDR   \\ 
\midrule
DiffSep~\cite{Diffsep}     &             & \checkmark  & 2.2436       &   0.6291     &   4.9435       \\
VisualVoice~\cite{visualvoice} & \checkmark    &       & 1.9708 & 0.7691 & 9.6810   \\
\midrule
\textbf{AVDiffuSS (Ours)}        & \checkmark   & \checkmark       & \textbf{2.5031}   & \textbf{0.8122} & \textbf{12.0282}  \\
\bottomrule
\end{tabular}%
}
\caption{Speech separation results on the VoxCeleb2 dataset. For all metrics, higher is better. A-V refers to audio-visual model, and Diff refers to diffusion-based model.}
\label{table1}
\end{table}

\newpara{Training objective.}
For an end-to-end training of predictor $D_\theta$ and generator $G_\phi$, we use a multi-task learning strategy, following~\cite{storm}. 
The equations below show the overall training process. The predictor $D_\theta$ is trained with L2 loss ($L_{pred}$) between the initial prediction $D_\theta(\mathbf{y})$ and the ground-truth $\mathbf{x}$,

\begin{equation}
    L_{pred} = \mathbb{E} \left[ \| \mathbf{x} - D_\theta(\mathbf{y})\|^2_2 \right]. 
\end{equation}

Our model in the reverse-diffusion stage is trained by utilizing a denoising score matching objective denoted as $L_{diff}$. This is a re-parameterized denoising score matching objective in~\cite{song2020score}, where $t$ is uniformly sampled within a range from minimal diffusion time $\tau_\epsilon$ to $T$.

\begin{equation}
    L_{diff} = \mathbb{E} \left[ \| \mathbf{s}_\phi(\mathbf{x_\tau}, \mathbf{y}, \tau) + \frac{\mathbf{z}}{\sigma_\tau} \|^2_2  \right].
\end{equation}

Two objectives are combined with weight values $\lambda_1$ and $\lambda_2$ for balanced training as follows:
\begin{equation}
    L = \lambda_1 * L_{pred} + \lambda_2 * L_{diff}.
\end{equation}

\begin{table}
\centering
{
\resizebox{\columnwidth}{!}{%
\begin{tabular}{lccccc} 
\toprule
Method   & A-V & Diff & PESQ   & ESTOI  & SI-SDR   \\ 
\midrule
DiffSep~\cite{Diffsep}     &              & \checkmark       & 2.0625       &   0.6826     &   5.3581       \\
VisualVoice~\cite{visualvoice} & \checkmark    &                 & 1.7778 & 0.7417 & 7.2270   \\
\midrule
\textbf{AVDiffuSS (Ours)}        & \checkmark    & \checkmark       & \textbf{2.6803} & \textbf{0.8769} & \textbf{13.6243}   \\
\bottomrule
\end{tabular}%
}
}
\caption{Speech separation results on the LRS3 dataset. Results are obtained by the models trained on VoxCeleb2 dataset.}
\label{table2}
\end{table}

\section{Experiments}
\label{sec:experiments}

We evaluate our model quantitatively using three established speech evaluation metrics, which are Perceptual Evaluation of Speech Quality (PESQ) \cite{pesq}, Scale-Invariant Signal-to-Distortion Ratio (SI-SDR)~\cite{sisdr}, Extended Short-Time Objective Intelligibility (ESTOI)~\cite{estoi}, and qualitatively through Mean Opinion Score (MOS). 

\subsection{Experimental Setup}
\label{ssec:expsetup}
\newpara{Datasets.}
VoxCeleb2~\cite{voxceleb2} dataset is a widely-used dataset for audio-visual tasks comprising more than 1 million utterances extracted from YouTube videos. This dataset consists of 5,994 identities in the training set, and 118 identities in the test set. Our model is trained on VoxCeleb2 train set, and 10 utterances in the test dataset are randomly chosen for validation. 
LRS3~\cite{afouras2018lrs3} dataset is another popular dataset for audio-visual speech recognition and speech separation. The dataset is made up of 4,004 videos for training and validation, and 412 videos for test set, which are from TED and TEDx videos. 

\newpara{Implementation details.}
We utilize NCSN++M for the U-net inside our model and the cross-attention mechanism is adopted from \cite{latentdif}. We follow the details regarding the diffusion process in~\cite{storm}. 
The input for the visual encoder is a sequence of face-cropped grayscale images resized to $112 \times 112$. 
Our model is updated with Adam optimizer~\cite{KingBa15} with an exponential moving average of network parameters with a decay of 0.999~\cite{song2020improved}. The initial learning rate is initiated to $10^{-4}$. The weight value $\lambda$ is set to 0.5. 
We use 4 RTX A5000 GPUs for training with effective batch size 16. We train our network for 15 epochs, which takes approximately 12 days. 

\newpara{Comparison methods.}
We compare our method with two publicly-available
state-of-the-art speech separation models.
DiffSep\footnote{https://github.com/fakufaku/diffusion-separation}~\cite{Diffsep} is an audio-only speech separation model extended from SGMSE+~\cite{sgmse+}, and
VisualVoice\footnote{https://github.com/facebookresearch/VisualVoice}~\cite{visualvoice} is an audio-visual speech separation model.
We train DiffSep on VoxCeleb2 dataset from scratch for 20 epochs for pair comparisons, as this results in approximately the same number of iterations as reported in~\cite{Diffsep}. We also utilize an official pre-trained VisualVoice model and generate a test set using the process introduced in~\cite{visualvoice}.
Note that every model is trained on VoxCeleb2 train set and tested on the test sets of VoxCeleb2 and LRS3.

\subsection{Experimental Results}
\label{ssec:exp_comparison}
\newpara{Quantitative results.}
To validate the effectiveness of our methods, we show the experimental results on VoxCeleb2 and LRS3 in ~\Tref{table1} and ~\Tref{table2}, respectively. 
In particular, DiffSep exhibits the lowest ESTOI and SI-SDR scores among the three models. This suggests that the utilization of the visual modality enables the models to effectively separate the target speech.
Moreover, when compared to VisualVoice, AVDifuSS shows a higher PESQ score. It indicates that our model generates natural-sounding speech.
To further simulate one-shot speech separation scenarios, we evaluate every model that is trained with VoxCeleb2 on LRS3 test set in~\Tref{table2}. This result shows the robustness of our model on a cross-dataset evaluation.
It is worth noting that our model is the first to leverage the diffusion mechanism in the task, and yet it significantly outperforms the existing state-of-the-art methods.

\newpara{Qualitative results.}
Generative models often receive low scores using conventional metrics due to the artifacts generated during the generation, even though they sound natural to the human ear.
Therefore, we conduct a human evaluation using MOS.
17 participants are each asked to assess 20 pairs of separated outputs on a scale of 1 to 5.
We normalize every sample to eliminate the amplitude bias in outputs of each model and the orders of the models are randomly assigned for every pair.
Criteria for the evaluation are (1) audio quality relative to the corresponding ground-truth, and (2) degree of separation. 
As some natural-sounding outcomes involve the other speaker's speech, it is impossible to evaluate the degree of separation without the ground truth samples. Therefore, the ground truth samples are not included in the MOS evaluation and are provided to the participants as standards for assessing separation capability. 
As shown in~\Tref{table:mos}, MOS for our model is significantly higher compared to previous works. These results demonstrate the ability of our approach to generate samples that sound clear and natural to human hearing, not to mention its intelligibility.

\begin{table}
\centering
\resizebox{0.7\linewidth}{!}{
\begin{tabular}{lccccc}
\toprule 
Method   & A-V & Diff &MOS  \\ 
\midrule
DiffSep~\cite{Diffsep}     &              & \checkmark       & 2.24 $\pm$ 0.11  \\
VisualVoice~\cite{visualvoice} & \checkmark    &        & 2.98 $\pm$ 0.10   \\
\midrule
\textbf{Ours}        & \checkmark   & \checkmark       & \textbf{4.44 $\pm$ 0.07} \\
\bottomrule
\end{tabular}
}
\caption{MOS comparison with 95\% confidence interval computed from the t-distribution. 17 participants rated 20 list of audios randomly selected from the results on VoxCeleb2.}
\label{table:mos}
\end{table}

\begin{table}[!t]
\centering
\resizebox{0.75\linewidth}{!}{
\begin{tabular}{lccc} 
\toprule 
Method & PESQ & ESTOI & SI-SDR  \\ 
\midrule
DiffSep~\cite{Diffsep}     & 1.8926     & 0.5116 & 0.4439      \\
VisualVoice~\cite{visualvoice} & 1.8063 & 0.7356 & 7.8063\\
\midrule
\textbf{Ours}              & \textbf{2.1171}     & \textbf{0.7566}   & \textbf{8.6177}    \\
\bottomrule
\end{tabular}
}
\caption{Speech separation results tested on the bottom 30\% samples sorted by SI-SDR results of VoxCeleb2 dataset.}
\label{table_bottom30}
\end{table}

\subsection{Discussions}
\label{ssec:ablation}

\newpara{Experimental results in difficult cases.}
In real-world scenarios such as a conversation of two speakers with similar timbre, it is difficult for audio-only speech separation models to accurately distinguish speech of each speaker.
Moreover, VisualVoice suffers especially from overlapping speeches of two male speakers with loud background noise due to the difficulty of clearly discriminating low-pitched voice with the babble noise. Thus we show the results from the hardest samples to prove the robustness of our diffusion-based audio-visual approach. 

Sorted by SI-SDR result of our model, we choose the bottom 30\% samples to demonstrate performance of each model in harsh conditions, which is disadvantageous to our method. 
As shown in Table~\ref{table_bottom30}, our model and VisualVoice are both able to isolate the whole intelligible speech even in hard cases by utilizing synchronization cues and facial characteristics. DiffSep cannot identify the target speaker accurately due to the lack of visual information, resulting in lower scores especially on SI-SDR.
On the other hand, diffusion-based approaches are able to generate outputs with high naturalness, indicated by PESQ metric. 

\begin{figure}[!t]
\centering
\includegraphics[width=1.0\linewidth]{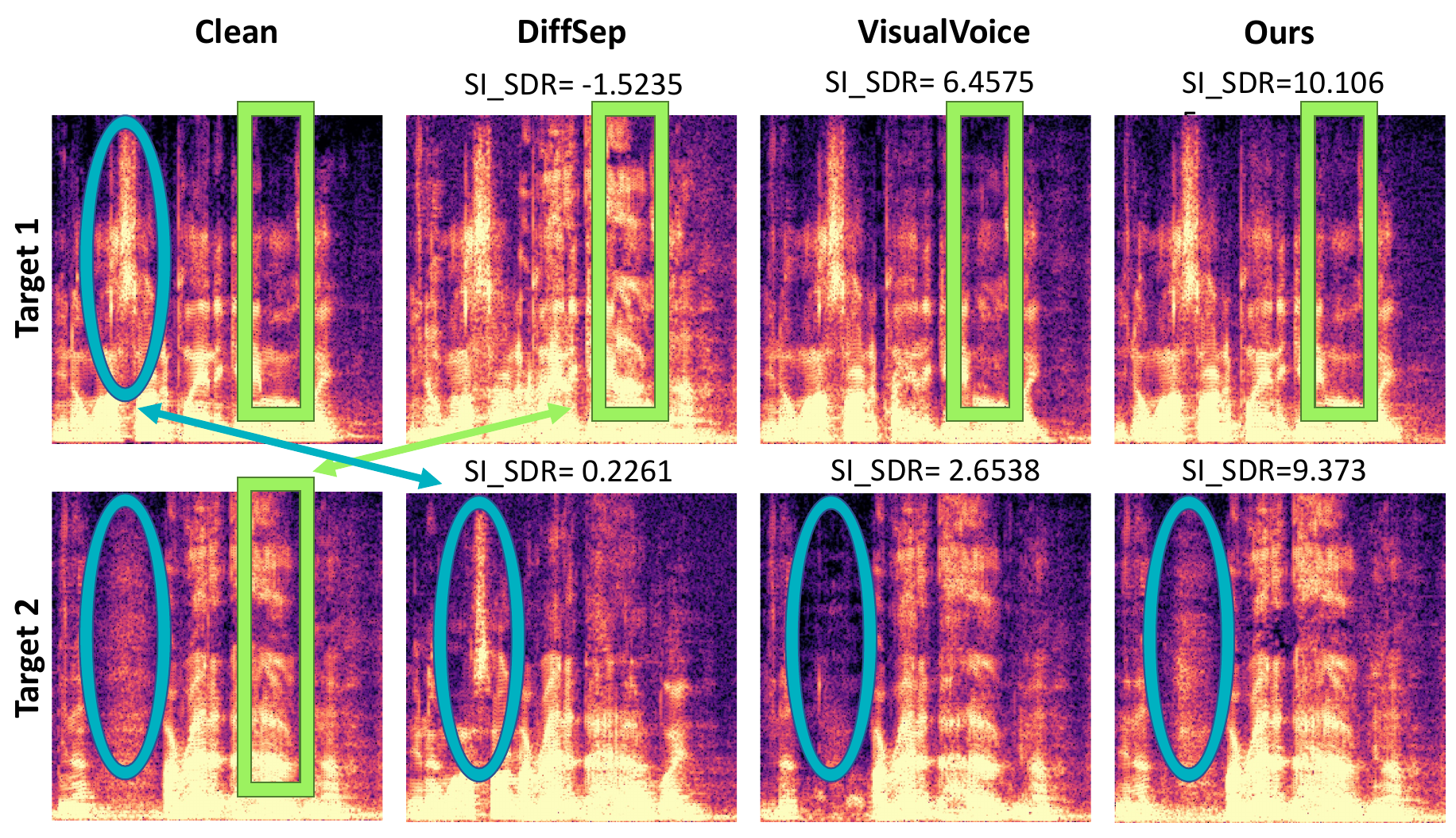}
\vspace{-2mm}
\caption{Spectrogram comparison of the outputs of DiffSep \cite{Diffsep}, VisualVoice~\cite{visualvoice} our model on the lowest 30\% random sample.}
\vspace{-2mm}
\label{fig3}

\end{figure}

Spectrograms of the separated outputs from each model are shown in Fig.~\ref{fig3}, including ground truth for a comparison. A pair of speech signals is randomly selected from the bottom 30\% results, and the mixture of speech is fed to each model to evaluate the three models. Boxes and circles with identical colors in each row represent regions which should be same with clean spectrogram.  
Arrows colored in green and blue denote that the non-target speaker's speech is included in the output of audio-only diffusion model, but not in our results. In VisualVoice, the details of the original speech is ignored, and the speech is over-denoised as shown in the spectrograms. This visualization demonstrates the ability of our model to generate realistic details, not to mention the accurate capturing of the spoken contents.

\begin{table}[!t]
\centering
\resizebox{0.8\linewidth}{!}{
\begin{tabular}{lccc} 
\toprule
Resolutions & PESQ & ESTOI & SI-SDR  \\ 
\midrule
32     & 2.4156    & 0.8000        & 11.5506\\
32, 64     & 2.4444 & 0.8028 & 11.6749    \\
\textbf{32, 64, 128 (Ours)}    & \textbf{2.5031}   & \textbf{0.8122} & \textbf{12.0282}   \\
\bottomrule
\end{tabular}
}
\caption{Ablation results on the VoxCeleb2 dataset depending on the feature resolutions to apply cross-attention choices.}
\label{table:abl}
\end{table}
\newpara{Effectiveness of cross-attention layers.}
Table~\ref{table:abl} shows the results of ablation on the resolution of the U-net features which the cross-attention modules are applied to. Starting from 256, the feature resolution of U-net layers are halved for four times from 256 to 32 during the downsampling path, and upsampled to recover the original feature resolution. Among those 8 layers, cross-attention modules are applied on the 6 layers with three smallest resolutions. 
Increasing the number of cross-attention layers brings a modest improvement in performance, but we do not add cross-attention to every layer in U-net due to memory constraints.

\section{Conclusion}
\label{sec:conclusion}
In this work, we present AVDiffuSS, an audio-visual speech separation framework based on the diffusion model. Our approach exploits visual cues to extract the target speaker's speech accurately, and the diffusion model to produce a highly natural-sounding output. 
We devise a task-specific feature fusion mechanism for integrating a target speaker's visual information into the diffusion-based speech separation framework.
The proposed model demonstrates state-of-the-art performance for audio-visual speech separation in terms of both naturalness and intelligibility.


\clearpage
\vfill\pagebreak

\bibliographystyle{IEEEbib}
\bibliography{shortstrings,refs}

\end{document}